\documentclass[12pt,nohyper,notoc]{JHEP}
\input epsf
\epsfverbosetrue
\usepackage{amsmath,euscript,array,amssymb,cite} 
\def\aD{{\dot\alpha}}
\def\bD{{\dot\beta}}
\def\M{{{\cal M}'}}
\def\N{{\cal N}}
\def\tr{{\rm tr}}
\def\trtwo{\tr^{}_2\,}
\def\Mbar{\bar{\cal M}}
\def\dalpha{{\dot\alpha}}
\def\dbeta{{\dot\beta}}

\def\wbar{\bar w}
\def\mubar{\bar\mu}
\def\vhiggsbar{\bar \vhiggs}
\def\sigmabar{\bar\sigma}

\def\mubar{\bar\mu}

\def\Tr{{\rm Tr}}
\def\sst{\scriptscriptstyle}
\def\det{{\rm det}}

\def\Dbarslash{\,\,{\raise.15ex\hbox{/}\mkern-12mu {\bar\D}}}
\def\Dslash{\,\,{\raise.15ex\hbox{/}\mkern-12mu \D}}
\def\delslash{\,\,{\raise.15ex\hbox{/}\mkern-9mu \partial}}
\def\delbarslash{\,\,{\raise.15ex\hbox{/}\mkern-9mu {\bar\partial}}}

\def\F{{\mathfrak F}}
\def\Z{{\EuScript Z}}

\def\hf{{\textstyle{1\over2}}}

\def\dalpha{{\dot\alpha}}
\def\alphabar{{\bar\alpha}}
\def\dbeta{{\dot\beta}}
\def\bp{{\boldsymbol p}}
\def\bzeta{{\boldsymbol\zeta}}
\def\N{{\cal N}}

\def\M{{\cal M}}

\def\Mbar{\bar{\M}}
\def\mubar{\bar{\mu}}

\def\homoa{\langle {\cal A}\rangle }
\def\barhomoa{\bar{\langle {\cal A}\rangle} }

\def\zbar{\bar{z}}

\def\sst{\scriptscriptstyle}

\def\F{{\cal F}}
\def\P{{\cal P}}
\def\A{{\cal A}}

\def\sigmabar{\bar\sigma}

\def\cl{{\,\rm cl}}
\def\lambdabar{\bar\lambda}
\def\R{{R}}
\def\psibar{\bar\psi}
\def\sqrtwo{\sqrt{2}\,}

\def\vhiggs{{\rm v}}

\def\vhiggsbar{\bar{\rm v}}

\def\zero{{\scriptscriptstyle(0)}}
\def\new{{\scriptscriptstyle\rm new}}

\def\uA{\,\lower 1.2ex\hbox{$\sim$}\mkern-13.5mu A}
\def\uX{\,\lower 1.2ex\hbox{$\sim$}\mkern-13.5mu X}
\def\uD{\,\lower 1.2ex\hbox{$\sim$}\mkern-13.5mu {\rm D}}

\def\uF{\,\lower 1.2ex\hbox{$\sim$}\mkern-13.5mu F}
\def\uW{\,\lower 1.2ex\hbox{$\sim$}\mkern-13.5mu W}
\def\uWbar{\,\lower 1.2ex\hbox{$\sim$}\mkern-13.5mu {\overline W}}
\def\uV{\,\lower 1.2ex\hbox{$\sim$}\mkern-13.5mu V}
\def\uv{\,\lower 1.0ex\hbox{$\scriptstyle\sim$}\mkern-11.0mu v}
\def\uPsi{\,\lower 1.2ex\hbox{$\sim$}\mkern-13.5mu \Psi}
\def\uPhi{\,\lower 1.2ex\hbox{$\sim$}\mkern-13.5mu \Phi}
\def\uchi{\,\lower 1.5ex\hbox{$\sim$}\mkern-13.5mu \chi}
\def\Psibar{\bar\Psi}
\def\uPsibar{\,\lower 1.2ex\hbox{$\sim$}\mkern-13.5mu \Psibar}
\def\upsi{\,\lower 1.5ex\hbox{$\sim$}\mkern-13.5mu \psi}
\def\psibar{\bar\psi}
\def\upsibar{\,\lower 1.5ex\hbox{$\sim$}\mkern-13.5mu \psibar}
\def\upsibarzero{\,\lower 1.5ex\hbox{$\sim$}\mkern-13.5mu \psibar^\zero}
\def\ulambda{\,\lower 1.2ex\hbox{$\sim$}\mkern-13.5mu \lambda}
\def\ulambdabar{\,\lower 1.2ex\hbox{$\sim$}\mkern-13.5mu \lambdabar}
\def\ulambdabarzero{\,\lower 1.2ex\hbox{$\sim$}\mkern-13.5mu \lambdabar^\zero}
\def\ulambdabarnew{\,\lower 1.2ex\hbox{$\sim$}\mkern-13.5mu \lambdabar^\new}
\def\D{{\cal D}}
\def\M{{\cal M}}
\def\N{{\cal N}}
\def\Dslash{\,\,{\raise.15ex\hbox{/}\mkern-12mu \D}}
\def\Dbarslash{\,\,{\raise.15ex\hbox{/}\mkern-12mu {\bar\D}}}
\def\delslash{\,\,{\raise.15ex\hbox{/}\mkern-9mu \partial}}
\def\delbarslash{\,\,{\raise.15ex\hbox{/}\mkern-9mu {\bar\partial}}}
\def\L{{\cal L}}
\def\hf{{\textstyle{1\over2}}}
\def\quarter{{\textstyle{1\over4}}}

\def\fourth{\quarter}

\def\uAcl{\,\lower 1.2ex\hbox{$\sim$}\mkern-13.5mu A^{}_{\cl}}
\def\uAbarcl{\,\lower 1.2ex\hbox{$\sim$}\mkern-13.5mu A_{\cl}^\dagger}

\def\Atot{{\cal A}_{\rm tot}}
\def\mubar{{\bar\mu}}

\def\Foneinst{{\cal F}_1}

\def\bigL{{\bf L}}
\def\Lambdabar{\bar\Lambda}
\def\Lambdatot{{\Lambda_{\rm tot}}}

\def\bigR{{\rm I}\!{\rm R}}

\def\Atot{{\A_{\rm tot}}}

\def\uA{\,\lower 1.2ex\hbox{$\sim$}\mkern-13.5mu A}
\def\bigL{{\bf L}}

\def\zero{{\scriptscriptstyle(0)}}
\def\calB{{\cal B}}

\newcommand{\EQ}[1]{\begin{equation} #1 \end{equation}}
\newcommand{\AL}[1]{\begin{subequations}\begin{align} #1 \end{align}\end{subequations}}
\newcommand{\SP}[1]{\begin{equation}\begin{split} #1 \end{split}\end{equation}}

\def\beqa{\begin{eqnarray}} 
\def\eeqa{\end{eqnarray}} 
\def\beq{\begin{equation}} 
\def\eeq{\end{equation}} 
\def\R{\mbox{\rm I\kern-.18em R}} 
\def\Rq{\R^4} 
\def\P{\mbox{\rm I\kern-.18em P}} 
\def\uno{\mbox{1 \kern-.59em {\rm l}}} 
\def\Z{{Z \kern-.45em Z}} 
\def\Q{{\kern .1em {\raise .47ex \hbox{$\scriptscriptstyle |$}} 
\kern -.35em {\rm Q}}} 
 
\def\Tr{\mbox{\rm Tr}} 
\def\Im{\mbox{\rm Im}} 
\def\cl{\mbox{\scriptstyle cl}} 
\def\tr{\mbox{\rm tr}} 
\def\p{\partial}

\def\ie{{\it i.e. }} 
 
\font\mybb=msbm10 at 12pt

\def\bb#1{\hbox{\mybb#1}} 
\def\Z {\bb{Z}}

\title{Exact Results in Noncommutative ${\cal N}=2$ Supersymmetric 
Gauge Theories}

\author{Timothy J.~Hollowood$^{a}$,
Valentin V.~Khoze$^b$ and Gabriele Travaglini$^b$\\
$^a$Department of Physics, University of Wales Swansea,
Swansea, SA2 8PP, UK\\
$^b$Department of Physics and IPPP, University of Durham,
Durham, DH1 3LE, UK\\
E-mail: {\tt t.hollowood@swan.ac.uk},
{\tt valya.khoze@durham.ac.uk}, {\tt gabriele.travaglini@durham.ac.uk}}

\abstract{We study the low-energy dynamics of noncommutative $\N=2$
supersymmetric $U(N)$ Yang-Mills theories in the Coulomb phase. Exact results
are derived for the leading terms in the derivative expansion of the Wilsonian
effective action. We find that in the infrared regime the $U(1)$ subgroup
decouples, and the remaining $SU(N)$ is described by the ordinary commutative
Seiberg-Witten solution. IR/UV mixing is present in the $U(1)$, but not in
$SU(N)$. Our analysis is based on explicit perturbative and multi-instanton 
calculations.} 


\preprint{{\tt hep-th/0102045}}

\begin{document}

\section{Introduction}\label{sec:S1}

Recently there has been a lot of interest in gauge theories on 
noncommutative spaces. One of the reasons is the natural appearance
of noncommutativity in the framework of string theory and D-branes
\cite{CDS,DHull,SWnc}.
Noncommutative gauge theories are also fascinating on their own right
mostly due to a new interplay between the infrared (IR) and the ultraviolet
(UV) degrees of freedom discovered in \cite{Minwalla}.
It is also known that this IR/UV mixing does not occur in 
$\N=4$ supersymmetric noncommutative gauge theories \cite{Matusis}.
This is supported by the $\N=4$ gauge/supergravity correspondence
discussed in \cite{Hashimoto:1999ut,Maldacena:1999mh}.

In this paper we analyse the leading terms in the derivative
expansion of the Wilsonian effective action for $\N=2$ supersymmetric
$U(N)$ gauge theories\footnote{This week
similar issues were addressed from a different perspective in
the interesting work \cite{Armoni:2001br}.}
on noncommutative space with 
$[x^\mu, x^\nu]=i\theta^{\mu\nu}.$
Specifically we will concentrate on the terms with at most two
space-time derivatives and/or not more than four fermions.
Such terms in the Wilsonian Lagrangian will be denoted
${\cal L}_{\rm eff}$.
We will demonstrate that for Wilsonian momentum scales $k^2$ below the
noncommutativity mass-scale, $k^2\ll M^2_{\sst NC}\sim \theta^{-1},$ the 
$U(1)$ degrees of freedom decouple from the $SU(N)$ fields 
and
\EQ{
{\cal L}_{\rm eff}^{\sst U(N)} (k) \ = \ 
{\cal L}_{\rm eff}^{\sst U(1)} (k) \ + \ 
{\cal L}_{\rm eff}^{\sst SU(N)} (k) \ .
\label{decrel}}
We will concentrate
on the Coulomb branch of the theory and parametrize the
vacuum expectation values (VEVs) of the adjoint scalar field via
\EQ{
\langle \varphi \rangle\ =\ {\rm diag}(\vhiggs_{\sst 1}, \ldots, \vhiggs_{\sst N}) \ . \label{vevs}}
Without loss of generality this matrix of VEVs can be chosen traceless,
$\sum_{u=1}^N \vhiggs_u =0,$
since the $U(1)$-part of the scalar VEV, 
$\langle A \rangle = {\rm V_{tr}}\, {\rm diag}(1,\ldots,1),$ 
breaks no symmetries
 and does not play any role in the dynamics of the theory. 
In fact, as noticed in \cite{Armoni:2001br}, 
the transformation ${\rm V_{tr}} \to {\rm V_{tr}} +  const$
is a symmetry of the theory and  
${\rm V_{tr}}$ is not a coordinate on the quantum moduli space. 

Since the noncommutative $U(1)$  decouples from $SU(N)$,
it can be analysed separately on its own right. Such
an analysis of the Wilsonian action
${\cal L}_{\rm eff}^{\sst U(1)} (k)$ was carried out
in the earlier work \cite{Khoze:2001sy} at the one-loop level, where it was found
that, due to the IR/UV mixing, the $U(1)$ 
$\N=2$ theory remains noncommutative even in the IR region, 
$k \ll M_{\sst NC},$ and arbitrarily weakly coupled
as $k^2\to 0,$ justifying the one-loop analysis. 
The approach of \cite{Khoze:2001sy} determines
the RG flow of the Wilsonian $U(1)$ coupling constant,
$g^{\prime \,2}_{\rm eff}(k),$ in such a way that 
\beqa
{1\over g^{\prime\, 2}_{\rm eff}(k)}  &\rightarrow&
{1 \over 8\pi^2} \log k^2 \ , \qquad {\rm as} \ 
k^2\to\infty \ ,
\\ 
{1\over g^{\prime\, 2}_{\rm eff}(k)} &\rightarrow&
{1 \over 8\pi^2} \log {1\over k^2} \ , \qquad {\rm as} \
k^2\to 0 \ .
\eeqa 
Thus, the noncommutative $U(1)$ theory is asymptotically free 
and weakly coupled in the UV region, and
it changes its behaviour to a screening regime at $k\sim M_{\sst NC}$,
and becomes arbitrarily weakly coupled in the IR. This running
of $g^{\prime 2}_{\rm eff}$
is strikingly different from the coupling of the ordinary $\N=2$
commutative $U(1)$ which is $k$-independent, {\it i.e.} does not run.
The corresponding 2-derivative Wilsonian action reads:
\EQ{
{\cal L}_{\rm eff}^{\sst U(1)} (k) \ = \ 
-{1  \over 2g^{\prime\, 2}_{\rm eff}(k) } 
 \ \Tr \left( F^{\sst U(1)}_{\mu \nu} \star  
F^{\sst U(1)}_{\mu \nu} \right)   
\ + \ \ldots
\ , 
\label{uoneres}}
where the dots stand for the $\N=2$ superpartners of the $U(1)$
gauge kinetic term, and the star-product is defined in the standard way as
\EQ{(\phi \star \chi) (x) \equiv \phi(x) e^{{i\over 2}\theta^{\mu\nu}
\stackrel{\leftarrow}{\partial_\mu}
\stackrel{\rightarrow}{\partial_\nu}}  \chi(x) \ . \label{stardef}}

Let us now discuss the $SU(N)$ degrees of freedom. The Higgs VEVs
\eqref{vevs} spontaneously break the gauge symmetry
$SU(N)\rightarrow U(1)^{N-1}$ by giving masses to the W-bosons
and their superpartners, $M_{W} \propto |\vhiggs_u-\vhiggs_v|,$ and leave
the $N-1$ fields in the Cartan subalgebra of $SU(N)$ massless.
At momentum scales $k < M_{W}$ the massive degrees of freedom 
will be integrated out leading to 
the Wilsonian action, ${\cal L}_{\rm eff}^{\sst [N-1]} (k),$
of the $N-1$ massless photons and their superpartners.
One of the principal results of this paper is the fact that 
${\cal L}_{\rm eff}^{\sst [N-1]}$ actually does not depend on $k$
for $k < M_{W}$.
It will turn out that the running $SU(N)$ coupling constant
will freeze at the momentum scale $k=M_{W}$ and the resulting
$(N-1)\times(N-1)$  matrix of coupling constants in the $U(1)^{N-1}$ low-energy
theory will not depend on $k$. Instead it will depend on the
VEVs $\vhiggs_{\sst 1}, \ldots \vhiggs_{\sst N}$ 
which set the values of the W-masses, where the freezing occurs. 
In other words, $\vhiggs_{\sst 1}, \ldots \vhiggs_{\sst N}$
will parametrize the vacuum moduli space of the $\N=2$ 
noncommutative $SU(N)$ theory similarly to the 
situation in the ordinary commutative $\N=2$ scenario
of Seiberg and Witten \cite{SW}.

This relation between the noncommutative and the commutative
$\N=2$ $SU(N)$ theories in the Coulomb branch is more than
just an analogy, it is an equivalence.\footnote{Our analysis
will be valid of course only for the leading term in the derivative expansion
of the low-energy effective action.} 
We will demonstrate below
that no IR/UV mixing occurs in the $SU(N)$ sector in perturbation
theory and non-perturbatively. This is unexpected for
noncommutative theories with $\N<4$ supersymmetries. 
The fact is that the IR/UV mixing in $U(N)$ occurs only in
the $U(1)$ part of the theory which decouples from the $SU(N)$
degrees of freedom and can be neglected in the IR. This  $U(1)$
theory is massless and can be characterized as being  
in the `noncommutative Coulomb phase'. 
The low-energy $U(1)^{N-1}$ theory ${\cal L}_{\rm eff}^{\sst [N-1]}$
can be formulated in superspace similarly to the ordinary theory
\cite{Ferrara,Terashima:2000xq}.
 Hence, it can be written in terms of
the holomorphic prepotential $\F(\Phi)$
\EQ{{\cal L}_{\rm eff}^{\sst [N-1]}\ =\ 
{\rm Im}\,{1\over4\pi}\,\Big[\,
\int d^4\theta\,{\partial\F(\Phi)\over\partial \Phi_I}\star\bar \Phi_I\ +\ 
{1\over2}\,\int d^2\theta\,{\partial^2\F(\Phi)\over\partial \Phi_I\,\partial \Phi_J}
\star W_I\star W_J\,\Big]\ .
\label{WilsF}}
Here $\Phi_I$ and $W_I$ are the $\N=1$ chiral $U(1)^{N-1}$
superfields containing
the $I$th massless Higgs boson and the $I$th photon field strength,
respectively, and $I,J=1,\ldots,N-1$. 
The prepotential $\F$ is a holomorphic
function of the $\N=1$ chiral superfields $\Phi_I$
and in general it can also depend on the
noncommutativity parameters $\theta^{\mu\nu}$. 
For the purposes of the low-energy theory we will set the
superfields $\Phi_I$ in the prepotential equal
to their VEVs $\vhiggs_I$. The matrix of the Wilsonian coupling
constants of the $U(1)^{N-1}$ theory is determined via 
\EQ{
{\partial^2\F(\vhiggs)\over\partial \vhiggs_I\,\partial\vhiggs_J} \ =\
\tau (\vhiggs)_{IJ} \ = \ {4\pi i \over g^2_{\rm eff}(\vhiggs)_{IJ}}
\ + \ {\vartheta_{\rm eff}(\vhiggs)_{IJ}\over 2 \pi} \ ,
\label{taudef}
} 
where $\vartheta_{\rm eff}$ is the effective theta-angle.

It is well-known \cite{Seiberg} that the prepotential is completely specified
by a perturbative one-loop contribution, and an infinite multi-instanton expansion
\EQ{
\F(\vhiggs,\theta)\ =\ \F_{1\hbox{-}{\rm loop}}(\vhiggs,\theta)\ +\
i \sum_{k=1}^\infty\,\F_k(\vhiggs,\theta) \ ,
\label{multiexp}}
where $\F_k$ denotes the contribution of the $k$-instanton sector.
In the rest of this paper we will calculate the
perturbative contribution $\F_{1\hbox{-}{\rm loop}},$ and deduce
all the multi-instanton
contributions to the prepotential from the field theory side.
 We will find that $\F$ does not
depend on $\theta$ and agrees precisely with the ordinary
commutative Seiberg-Witten prepotential.

The rest of the paper is organized as follows.
In Section 2 we study the $N^2\times N^2$
matrix of Wilsonian coupling constants
of the noncommutative
$U(N)$ theory at one-loop level using the background field method.
We find that after decoupling of massive degrees of freedom
this matrix factorizes in the IR region,
\EQ{
{1\over g^2 (k)}_{\sst [N^2]\times[N^2]} \ \rightarrow \
{1\over g_{\rm eff}^2 (k)}_{\sst[N]\times[N]} \ = \ 
{1\over g^{\prime\, 2}_{\rm eff}(k)} \ \oplus \ 
{1\over g_{\rm eff}^2 (\vhiggs)}_{\sst[N-1]\times[N-1]} \ 
\label{wcfac}
} 
which corresponds to the perturbative\footnote{It will follow from 
the analysis
in Section 3 that this decoupling is also
respected by instantons.} decoupling of 
$U(N)\rightarrow U(1) \times SU(N).$ The IR/UV mixing effects will be
present in the $U(1)$ coupling and absent in the $SU(N)$ coupling constant.
The latter
will be frozen at the W-mass scale $\vhiggs$, and its dependence on
$\vhiggs$ will be 
exactly the same as in the commutative $SU(N)$ theory, hence it 
will match precisely with the commutative Seiberg-Witten prepotential
$\F_{1\hbox{-}{\rm loop}}.$

In Section 3 we consider instanton and anti-instanton contributions to the
low-energy effective action in the noncommutative $U(N)$ gauge theory.
We give a general argument that all multi-instanton contributions
to the prepotential in the noncommutative case do not depend
on the noncommutativity parameter and agree with the ordinary
commutative contributions.
In Section 4 this general multi-instanton argument is explicitly
verified with a detailed one-instanton calculation.

\centerline{\it Note on Conventions}

{\parindent=0 pt
We introduce anti-hermitian generators of $U(N)$  as 
$t^A$, $A =(0, a)$, where $a = 1, \ldots , N^2-1$ 
labels the $SU(N)$ generators, and $t^0 = ( 1 / i \sqrt{2N}) \uno_{N}$. 
Then 
\EQ{
\Tr (t^A t^B ) = - {\delta^{AB} \over 2} 
\ \ .
}
The generators satisfy 
\beqa
[t^A, t^B] &=& f^{ABC} t^C
\ , 
\\ 
\{ t^A, t^B \} &=& - { \delta^{AB} \over N}- i d^{ABC} t^C
\ \ .
\eeqa
$f^{ABC}$ ($d^{ABC}$) is completely antisymmetric (symmetric) in its indices;    
$f^{abc}$, $d^{abc}$ are the same as in $SU(N)$, and 
$f^{0bc}= 0$, $d^{0BC} = \sqrt{2\over N} \delta_{BC}$, $\delta^{00a}= 0$, 
$d^{000} = \sqrt{2\over N}$.

Given an arbitrary four-vector, we will also use the notation 
$\tilde{k}_\mu \equiv  \theta_{\mu \nu}k_\nu$.

The Euclidean  $\sigma_\mu$  and $\bar{\sigma}_\mu$ matrices are defined as
$\sigma_\mu=(\uno_{2\times 2},i\sigma^m)$ and  
$\bar{\sigma}_\mu=(\uno_{2\times 2},-i\sigma^m)$ where $\sigma^m$   
are the three Pauli matrices. 
We will also use  
$\sigma_{\mu \nu}= {1\over 2} (\sigma_\mu \bar{\sigma}_\nu-\sigma_\nu \bar{\sigma}_\mu)=  
i \eta^{a}_{\mu \nu}\sigma^a$, and  
$\bar{\sigma}_{\mu \nu}=  
{1\over 2} (\bar{\sigma}_\mu \sigma_\nu-\bar{\sigma}_\nu \sigma_\mu)=  
i \bar{\eta}^{a}_{\mu \nu}\sigma^a$, where $\eta^{a}_{\mu \nu}$ and  
$\bar{\eta}^{a}_{ \mu \nu}$ 
are the 't Hooft symbols \cite{'tHooft:1976}.

}

\bigskip

\section{One-loop calculation of the effective action}

In this Section we will apply the background field perturbation
theory to noncommutative $U(N)$. Our discussion here follows closely
the formalism introduced in 
\cite{Khoze:2001sy}, to which we refer the reader for further details.
The gauge field $A_\mu$ is decomposed into a background field
$B_\mu$ and a fluctuating quantum field $N_\mu$, 
\EQ{
\label{fdec}
A_\mu= B_\mu + N_\mu 
\ \ , 
}
where $N_\mu$ is a highly virtual field 
with momenta above the Wilsonian scale.  
The background field is slowly varying, 
but fully noncommutative. The 
effective action $S_{\rm eff}(B)$  is obtained by functionally
integrating over the fluctuating fields. 
Noncommutative gauge-invariance   
constrains the interactions which can be generated  
in this procedure. Therefore, the 
effective action will always contain the kinetic term
\EQ{
S_{\rm eff} [ B ] \ni -{1  \over 2g^2_{\rm eff}} 
\int d^{4} x \ \Tr \left( F_{\mu \nu}^{(B)}\star  F_{\mu \nu}^{(B)} \right)   
\ \ . 
}
The multiplicative coefficient on the right hand side is identified
with the Wilsonian coupling constant at the corresponding momentum scale.
In order to determine $g_{\rm eff}$ it is sufficient   
to consider  the kinetic term $(\p_\mu B_\nu - \p_\nu B_\mu)^2$.  
In the effective Lagrangian, this term 
becomes
\EQ{
2 \int {d^4 k \over (2\pi)^4} B_\mu^{A} (k) B_\nu^{B} (-k)   
\ \Pi_{\mu \nu}^{AB} 
\ \ . 
\label{wptdef} 
}
Equation (\ref{wptdef}) defines the {\it Wilsonian polarization tensor} $\Pi_{\mu \nu}^{AB} (k)$, 
which in the effective theory replaces the tree level transverse tensor 
$(k^2 \delta_{\mu\nu}-k_\mu k_\nu )$.
On general grounds, $\Pi_{\mu \nu}^{AB} (k)$ has the structure
\EQ{
\Pi_{\mu \nu}^{AB} (k) =  
\Pi_1^{AB} (k^2, \tilde{k}^2) (k^2 \delta_{\mu\nu}-k_\mu k_\nu ) +  
\Pi_2^{AB} (k^2, \tilde{k}^2) {\tilde{k}_\mu\tilde{k}_\nu \over \tilde{k}^4} 
\ \ . 
\label{pimunu}
}
Here $\Pi_1^{AB}(k^2, \tilde{k}^2)$ determines the matrix of the 
Wilsonian couplings,
\EQ{
\left[{1\over g_{\rm eff}^2 (k)}\right]^{AB} \ = \ 
{\delta^{AB}\over g_{\rm micro}^2 } \ + \
4 \Pi_1^{AB} (k^2, \tilde{k}^2) \ .
}
The term in \eqref{pimunu} 
proportional to $\tilde{k}_\mu\tilde{k}_\nu /\tilde{k}^4 $ 
would not appear in 
ordinary commutative theories. It is transverse and 
has derivative
dimension $-2$; therefore it is of leading order
compared to the standard gauge-kinetic term
(which has  derivative dimension $+2$), and, most importantly, 
leads to an infrared singular behaviour.
In  \cite{Khoze:2001sy} it was shown 
that $\Pi_2$ vanishes
for all supersymmetric noncommutative $U(1)$ gauge theories
(unbroken and softly broken), as was first discussed   
in \cite{Matusis}. $\Pi_2$ is an intrinsically noncommutative object  
and arises only from nonplanar diagrams, whereas 
$\Pi_1$ receives contribution from planar 
as well as from  nonplanar diagrams.

The action functional which describes the dynamics of  
a spin-$j$ noncommutative field in the representation  
{\bf r} of the gauge group in the background of $B_\mu$ has the general
form \cite{Khoze:2001sy,Peskin}
\beqa 
S [ \phi ] &=&  
-\int d^{4} x \  \phi_{m,a} \star  
\left ( - D^2 (B)\delta_{mn}\delta^{ab} + 2i (F_{\mu \nu}^B)^{ab}  
\hf J^{\mu \nu}_{mn}   
\right)\star \phi_{n,b} 
\cr 
&\equiv&-\int d^{4} x \   
\phi_{m,a} \star [ \Delta_{j,{\bf r}}]_{mn}^{ab}\star \phi_{n,b} 
\ \ .  
\eeqa 
Here $a,b$ are indices of the  representation {\bf r} of noncommutative $U(N)$,  $F^{ab}\equiv \sum_{A=1}^{N^2} F^A t^A_{ab}$, and $m,n$ are spin indices and  
$J^{\mu \nu}_{mn}$ are  
the generators of the euclidean Lorentz group appropriate  
for the spin of $\phi$: 
\beqa 
J &=& 0 \qquad \qquad \qquad \qquad {\rm for\ spin\ 0 \ fields}, 
\\ \nonumber 
J^{\mu \nu}_{\rho \sigma}&=&i(\delta^\mu_\rho \delta^\nu_\sigma - 
\delta^\nu_\rho\delta^\mu_\sigma) \qquad {\rm for\ 4\hbox{-}vectors}, 
\\ \nonumber 
[J^{\mu \nu}]_{\alpha}^{\ \beta} &=& 
i  \hf [\sigma^{\mu \nu}]_{\alpha}^{\ \beta}
\qquad \qquad {\rm for\ Weyl\ fermions}\ \ . 
\eeqa 

At the one-loop level, the effective action is given by \cite{Khoze:2001sy}
\EQ{
\label{noncera}
S_{\rm eff} [B] =  
-{1  \over 2g^2} 
\int d^{4} x \ \Tr F_{\mu \nu}^B\star  F_{\mu \nu}^B - 
\sum_{j,{\bf r}} \alpha_{j} \log {\rm det}_{\star} \Delta_{j, {\bf r}}  
\ \ , 
}
where the sum is extended to all fields in the theory, including ghosts and gauge fields.  
$\alpha_{j}$ is equal to $+1$ ($-1$) for ghost (scalar) fields
and to $+1/2$ ($-1/2$) for 
Weyl fermions (gauge fields). 
Functional star-determinants are computed by 
\SP{\log {\rm det}_{\star} \Delta_{j, {\bf r}} \equiv&
\log {\rm det}_{\star} (-\partial^2 + {\cal K}(B)_{j, {\bf r}})\\
&= \log {\rm det}_{\star} (-\partial^2) + 
{\rm tr}_{\star} \log (1+(-\partial^2)^{-1}{\cal K}(B)_{j, {\bf r}}) \ .
\label{logkdef}
}
The first term on the second line of \eqref{logkdef} contributes only to the vacuum loops
and will be dropped in the following.
The second term on the last line of \eqref{logkdef} has an expansion
in terms of Feynman diagrams.   

 
\subsection{Feynman rules}
Since our main target is the computation of the effective action 
for ${\cal N}=2$ noncommutative Super Yang-Mills, we restrict our attention 
to fields transforming according to the adjoint representation of the group $U(N)$.
We start off our analysis considering the case of vanishing vacuum expectation value for 
scalar fields, postponing the discussion of spontaneously broken theories 
to  Section \ref{sbrot}.

As in \cite{Khoze:2001sy}, we rewrite $\Delta_{j, {\bf r}}$  
acting on adjoint fields as   
\SP{
\Delta_{j, {\bf \sst G}}\star \phi &\equiv 
-\partial^2 \phi + {\cal K}(B)_{j, {\bf \sst G}}\star \phi \\
&=-\partial^2 \phi -  \left[ (\p_\mu B_\mu ) , \phi \right]_\star -  
2 \left[ B_\mu \p_\mu , \phi \right]_{\star}-  
\left[ B_\mu ,  \left[ B_\mu , \phi \right]_\star  \right]_{\star}  
+ 2i \left( \hf J^{\mu \nu}  
\left[ F_{\mu \nu}^B , \phi \right]_{\star} 
\right) 
\ \ .  
\label{deladj}
}
The main difference with respect to the $U(1)$ case considered
in \cite{Khoze:2001sy} is that now
\EQ{
[\phi_1 , \phi_2]_{\star} = \left(-{i\over 2} [\phi_1^A , \phi_2^B]_{\star} d^{ABC} + {1\over 2}\{\phi_1^A , \phi_2^B\}_{\star} f^{ABC}\right)t^C
\ \ .
}
The Taylor expansion of the logarithm in \eqref{logkdef}
will involve the Feynman diagrams made from the three interaction vertices.
The first one,  the $\phi$-$B$-$\phi$ vertex, 
follows from the second and the third term
on the second line in \eqref{deladj}, 
\vfil\eject
\beqa  
\label{ncsqued1} 
- 2 \Tr &&\int d^4x \ \bar{\phi} \star  
\left[ (\p_\mu B_\mu ) + 2 B_\mu \p_\mu , \phi \right]_{\star} \ =\ 
\int {d^4 p' \over (2\pi)^4}   
{d^4 q \over (2\pi)^4}   
{d^4 p \over (2\pi)^4} \  
(2\pi)^4 \delta^{(4)} (p+q-p')
\cr \cr  
&& 
\bar{\phi}^A(p')  B_\mu^B (q) \phi^C (p)  \,
\left[ i (2p+q)_{\mu} (- d^{ABC}\sin {q \tilde{p} \over 2} 
+ f^{ABC}\cos {q \tilde{p} \over 2})\right] 
\ \ . 
\eeqa 
The second vertex $\phi$-$B$-$B$-$\phi$
follows from the fourth term
on the second line in \eqref{deladj},
\beqa 
\label{ncsqued2} 
- 2\Tr &&\int d^4x \ \bar{\phi} \star  
\left[ B_\mu , \left[ B_\mu , \phi \right]_\star \right]_{\star} 
 \ = \cr \cr 
&&\int {d^4 p' \over (2\pi)^4}   
{d^4 q_1 \over (2\pi)^4}  {d^4 q_2 \over (2\pi)^4}   
{d^4 p \over (2\pi)^4} \ 
(2\pi)^4 \delta^{(4)} (p+q_1+ q_2-p')\ 
\bar{\phi}^A(p')  B^B_\mu (q_1)  B^C_\nu (q_2)\phi^D (p)   \delta_{\mu \nu} 
\cr \cr
&&( -d^{BHA}\sin {p' \tilde{q_1}\over 2}  +  
f^{BHA}\cos {p' \tilde{q_1}\over 2} )
\,
(d^{CDH}\sin {q_2 \tilde{p'}\over 2} + 
f^{CDH}\cos {q_2 \tilde{p}\over 2 }) 
\ \ , 
\eeqa 
and finally the third vertex follows from the last term
on the second line in \eqref{deladj}: 
\beqa  
\label{ncsqued3} 
2i\Tr &&\int d^4x \ \bar{\phi} J^{\mu \nu} \star  
\left[ \p_\mu B_\nu - \p_\nu B_\mu , \phi \right]_{\star} \ = \ 
\int {d^4 p' \over (2\pi)^4}   
{d^4 q \over (2\pi)^4}   
{d^4 p \over (2\pi)^4} \  
(2\pi)^4 \delta^{(4)} (p+q-p')\  
\cr \cr 
&&\bar{\phi}^A(p')  J^{\mu \nu} B_B^\nu (q) \phi^C (p)
\,\left[ 2q_\mu (-d^{ABC}\sin { q\tilde{p}\over 2} + 
f^{ABC}\cos { q \tilde{p} \over 2}
)\right]   
\ \ . 
\eeqa 
The first two vertices (\ref{ncsqued1}) and (\ref{ncsqued2}) 
are the standard Feynman vertices for  
noncommutative electrodynamics with an adjoint scalar field,
and the third expression
\eqref{ncsqued3} is the so-called $J$-vertex, which arises in 
the background field method \cite{Peskin}.

\subsection{Planar and nonplanar contributions to the effective action}

Expanding the logarithm in \eqref{logkdef} to the second order
in the background fields $B_\mu$ gives the 
Feynman graphs shown in Figures 1, 2 and 3, in which   the 
$J$-vertices are  depicted by a cross.
Dimensional regularization is understood in all the integrals,
and UV-divergences are removed with the supersymmetry-preserving
$\overline{\rm DR}$-scheme \cite{Siegel}.
 
The first Feynman graph (shown in Figure 1) gives a contribution 
which reads   
\EQ{
\label{first} 
-{1 \over 2} \int {d^4 k\over (2\pi)^4} B_\mu (k) B_\nu (-k)    
\int {d^D p \over (2\pi)^D}  
\Tr { -(2p + k)_\mu (2p + k)_\nu M^{AB}(k, p)
\over 
p^2 (p+k)^2} 
\ \ , 
}
where we have introduced the tensor
\EQ{
\label{emme}
M^{AB}(k, p) = 
(-d \sin {k \tilde{p}\over 2}+ f \cos {k \tilde{p}\over 2})^{ALM}
(d \sin {k \tilde{p}\over 2}+ f \cos {k \tilde{p}\over 2})^{BML}
\ \ .
}
The second diagram, shown in Figure 2, gives 
\EQ{
\label{second} 
 \int {d^4 k\over (2\pi)^4} B_\mu^{A} (k) B_\nu^{B} (-k)    
\int {d^D p \over (2\pi)^D}  
\Tr {-\delta_{\mu \nu}\ M^{AB}(k, p)\over p^2} 
\ \ . 
}

\vfil\eject

\epsfxsize=18cm
\centerline{\epsffile{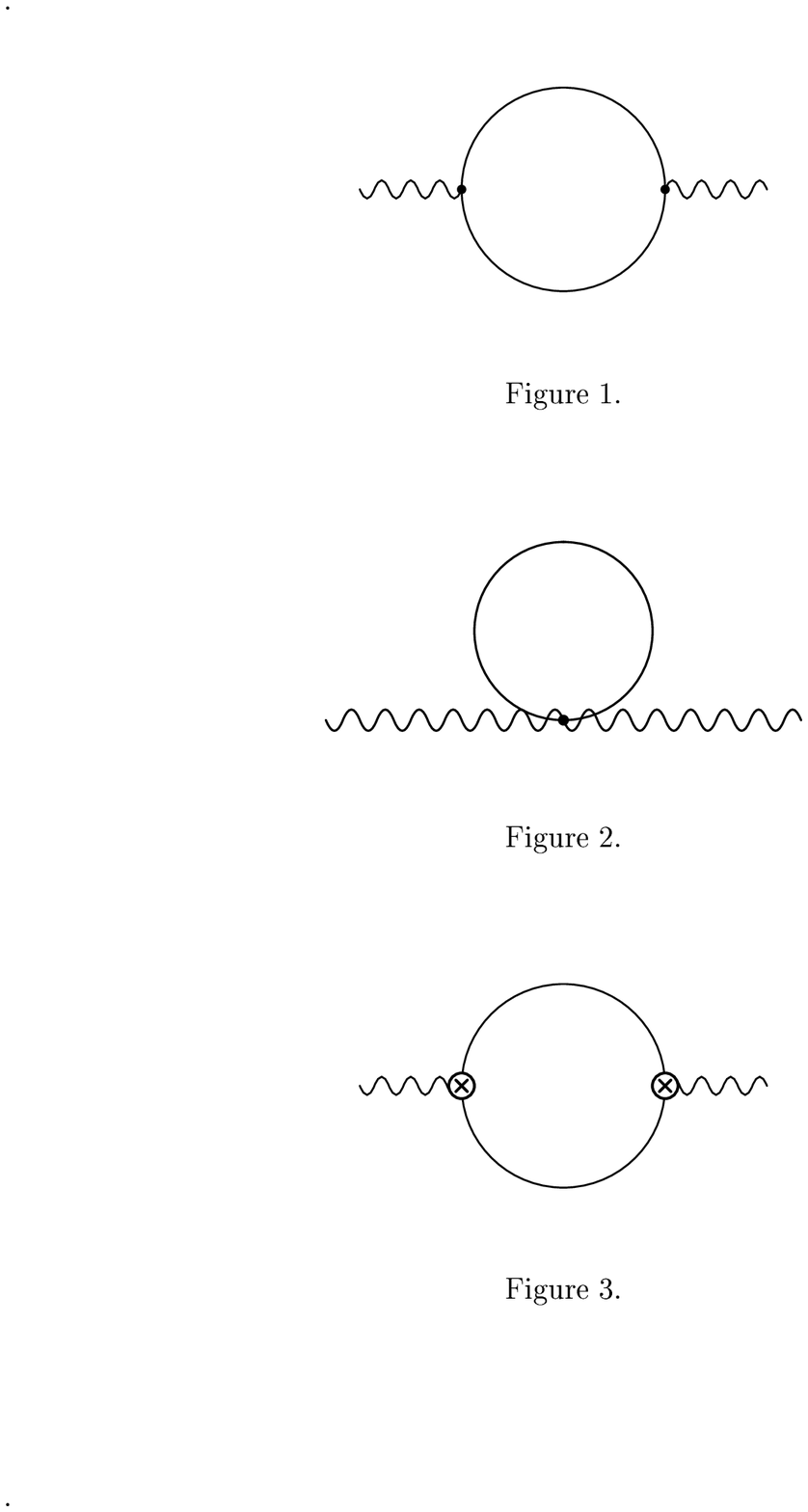}}

\vfil\eject

\noindent
In (\ref{first}), (\ref{second}) the trace is over spin indices, and  
its effect leads to a multiplicative factor of
\EQ{
\Tr \uno_{j} \equiv d(j) 
\ \ , 
}
where $d(j)$ is the number of spin component of the field $\phi$, 
\EQ{
d(j) \equiv \quad 1 \quad
{\rm for\,scalars,}\qquad 2 \quad {\rm for\, Weyl\, fermions,}
\qquad 4 \quad{\rm for\, vectors}. }
It is worth remarking that in supersymmetric theories the  cancellation
between bosonic and fermionic degrees of freedom enters the game via the 
identity
\EQ{
\label{susyzero} 
\sum_{j} \alpha_j d(j) = 0 
\ \ ,
} 
which holds for any representation of the gauge group. This in turn implies that the first and the second diagram separately vanish in any supersymmetric theory, even in presence of (supersymmetry preserving) spontaneous symmetry breaking.

We now move on to the   third amplitude, which  is depicted in Figure 3 
and gives  
\EQ{
\label{third} 
-{1\over 2}  \int {d^4 k\over (2\pi)^4} B_\mu^A (k) B_\nu^B (-k)    
\int {d^D p \over (2\pi)^D}  
\Tr {-4 J^{\mu \rho} J^{ \nu \lambda}  k_{\lambda} k_{\rho}  
 M^{AB}(k, p)\ \over p^2 (p+k)^2} 
\ \ , 
} 
where in the spin $j$ representation  
\EQ{
\Tr (J^{\mu \rho} J^{ \nu \lambda})_{j} = C(j) (\delta^{\mu \nu} \delta^{\rho \lambda} -  
\delta^{\mu \lambda} \delta^{\nu \rho}) 
\ \ ,  
}
\EQ{
C(j) \equiv \quad 0 \quad
{\rm for\,scalars,}\qquad \hf \quad {\rm for\, Weyl\, fermions,}
\qquad 2 \quad{\rm for\, vectors}. }

To proceed further on, we rewrite  (\ref{emme}) using the relations 
\cite{Bonora:2000ga}
\beqa
\label{tracce}
f^{ALM} f^{BML} &=& -N c_A \delta_{AB} \ \ , 
\nonumber \\
d^{ALM} d^{BML} &=& N d_A \delta_{AB} \ \ , 
\nonumber \\
f^{ALM} d^{BML} &=& 0 \ \ , 
\eeqa
where $c_A = 1 - \delta_{0A}$ and $d_A = 2-c_A$. 
This way (\ref{emme}) collapses to 
\EQ{
\label{emme2}
M^{AB}(k,p) = - N \ \delta^{AB} (1-\delta_{0A}\cos k\tilde{p})
\ \ .
}
The two terms on the right hand side of Eq. (\ref{emme2}) respectively select 
the planar and the nonplanar contribution, the latter  
explicitly depending on the noncommutativity parameter.
Using  \eqref{emme2}
we can recast the $U(N)$ polarization tensor as
\vfil\eject  
\beqa
[\Pi_{\mu \nu}^{AB} ]^{planar} [U(N)] &=& 
N \ \delta^{AB} \Pi_{\mu \nu}^{planar} [U(1)]
\ \ ,
\cr \cr
[ \Pi_{\mu \nu}^{AB} ]^{np} [U(N)] &=& 
N \ \delta^{A0} \delta^{B0}\Pi_{\mu \nu}^{np} [U(1)]
\ \ .
\label{pis} 
\eeqa 
Here 
$[\Pi_{\mu \nu}^{AB}]^{planar} [U(1)]$ 
and $\Pi_{\mu \nu}^{np} [U(1)]$ are respectively the planar and 
nonplanar contributions to the polarization tensor for gauge group $U(1)$, and 
have been calculated in \cite{Khoze:2001sy}. 
In particular $\Pi_{\mu \nu}^{np} [U(1)]$ contains the IR/UV mixing
terms characteristic to noncommutative $U(1)$ theories as shown in
\cite{Khoze:2001sy}.
Equations (\ref{pis})
remarkably show the decoupling of the $U(1)$ component 
associated with the generator $t^0\propto \uno$, as well as 
the absence of nonplanar contributions for the $SU(N)$ fields in the
effective action,%
\footnote{A similar decoupling between $U(1)$ and $SU(N)$ components 
was observed 
in \cite{Armoni} for the one-loop gluon propagator in 
noncommutative QCD with $N$ colours.} 
\EQ{
{1\over g_{\rm eff}^2 (k)}_{\sst[N^2]\times[N^2]} \ = \ 
{1\over g^{\prime\, 2}_{\rm eff}(k)} \ \oplus \ 
{1\over g_{\rm eff}^2 (k)}_{\sst[N^2-1]\times[N^2-1]} \ 
\label{wcfactwo}
} 
We now move on to consider the  
spontaneously broken case. 

\subsection{Spontaneously broken theories}
\label{sbrot}
In this subsection we follow the  analysis of \cite{Dorey:1997ij}
and focus only on  ${\cal N}=2$ noncommutative Super Yang-Mills theories in the Coulomb phase. 
They can be conveniently described as the dimensional reduction of 
${\cal N}=1$ Super Yang-Mills in 6 space-time dimensions down to 4 dimensions
\cite{Brink:1977bc}. More precisely, 
we extend the $U(N)$ gauge field $A_\mu$ 
to a 6-dimensional vector field incorporating the
two real scalars of pure ${\cal N}=2$ Super Yang-Mills:  
\def\sixd{{\sst\rm6D}}
\begin{equation}
A_\mu^\sixd\ =\ \big(\, A_1\,,\, A_2\,,\, A_3\,,\,
A_4\,,\, A_5\equiv\varphi_1\,,\, A_6\equiv\varphi_2\,\big)
\ \ .
\label{vsixdef}
\end{equation}
Assuming that all relevant field configurations are independent of 
the final two compactified spatial directions,
then the 4-dimensional ${\cal N}=2$ $U(N)$ Lagrangian  is known to
be simply that of 6-dimensional ${\cal N}=1$ Super Yang-Mills theory. 
At the one-loop level we focus on terms quadratic in the fluctuating
fields, and replace (\ref{fdec}) by 
\begin{equation}
\begin{array}{l l l}
A_\mu^\sixd  \ =\   &   &  B_\mu + N_\mu  \  \  \ , \ \ \ \mu  = 1, \ldots , 4 \ \ , 
\\
A_5^\sixd  \ = \     &{\rm v}^{(1)}\ + &  B_5 + N_5  \ \ ,    
\\ 
A_6^\sixd  \ = \     &{\rm v}^{(2)} \ + &  B_6 + N_6  \ \ .  
\label{deviation}
\end{array}
\end{equation}
Here ${\rm v}^{(1)} + i {\rm v}^{(2)}\equiv {\rm v}$ 
is the (constant) vacuum expectation value 
of the complex scalar field $\varphi$ of the 4-dimensional ${\cal N}=2$ supersymmetry 
which minimizes the potential 
$\Tr ([ \varphi , \varphi^\dagger ]_\star)^2$. 
As usual, ${\rm v}$ can  be expanded in terms of the Cartan generators $H_u$, $u=1, \ldots , N$
of the gauge group $G$ as 
\begin{equation}
\label{decartan}
{\rm v} = \sum_{u=1}^{N} {\rm v}_u H_u 
\ \ .
\end{equation}
The Higgs mechanism breaks the gauge symmetry to that of the Cartan subalgebra $H$, \ie\
$U(1)^N$. 
However,  notice that a noncommutative $U(1)$ theory is not spontaneously 
broken when its scalar acquires 
a nonvanishing vacuum expectation value $\varphi_0$ \cite{Armoni:2001br},  
as it follows directly from  observing  that 
$\Omega \varphi_0 \Omega^{-1} = \varphi_0$, where $\Omega \in U(1)_{\star}$. 
To see how this circumstance affects the full $U(N)$ theory, 
let us now move on to the calculation of the effective action. 
In the one-loop approximation, the only difference with the unbroken case 
comes from the new terms which appear 
in the generalized `kinetic' operator  (\ref{deladj}) as a consequence of expanding the 
6-dimensional gauge field as in (\ref{deviation}) with ${\rm v} \neq 0$. 
Without loss of generality we can set ${\rm v}^{(2)} = 0$; 
it is then immediately realized that, 
under the hypothesis of independence of the compactified dimensions, the 
relevant kinetic operators 
are replaced by 
\EQ{
\label{newdelta}
\Delta_{j, {\bf \sst G}}\star \phi \longrightarrow 
\Delta_{j, {\bf \sst G}}\star \phi - [{\rm v}, [{\rm v} , \phi]] - 
2 [{\rm v} , [B_5 , \phi]_{\star}]_\star
\ \ .
}
The last  term  in the right hand side of (\ref{newdelta}) corresponds to a new interaction vertex.
However, it is very easy to convince oneself that 
the corresponding  new contributions to the one-loop expansion of the logarithm 
in \eqref{logkdef} 
separately vanish as a consequence of supersymmetry, (\ref{susyzero}),  and of the 
property $\Tr \ [J^{\mu \nu} ] = 0$ of the Lorentz generators. 

The first term in the right hand side of (\ref{newdelta}) corresponds to a  mass term 
in the tree level action, 
\EQ{
\int dx \ \Tr ( \bar{\phi}\   [\bar{{\rm v}}, \ [ {\rm v}, \phi]\ ]) \equiv 
\int dx \ \bar{\phi}^A   {\cal M}^2_{AB}\phi^B    
\ \ , 
}
where the trace is in the group space.
Using for the generators of the $U(N)$ algebra a basis
$\{ H^u \ , E^{uv}_{\pm} (u>v)\}$, 
with $H^u_{AB}= \delta^{u}_{A}\delta^{u}_{B}$, and decomposing accordingly the fields
as $\sum_{u=1}^{N}\phi_{u}H^{u} + \phi_{uv}^{\pm}E_{\pm}^{uv}$, 
it is immediately seen that the Higgs mechanism 
gives the  $\phi_{uv}^{\pm}$ components masses proportional to the differences 
$|\vhiggs_u - \vhiggs_v|$. No dependence on the `center of mass' coordinate $\sum_{u=1}^{N}\vhiggs_{u}$
appears, which thus does not influence physics. 
Therefore, in the low-energy effective action $N$ massless supermultiplets are expected, 
but only $N-1$ moduli.

To efficiently perform perturbative expansions, 
it is convenient to define an  operator  ${\cal G}$ as 
the inverse in momentum space of the new tree level kinetic term, \ie\ 
${\cal G}^{AB}  \equiv [(-\partial^2 +{\cal M}^2)^{-1}]^{AB}$;  
\def\calG{{\cal G}}
for example, when the gauge group is $U(2)$ we can without loss of generality 
set ${\rm v}^a \propto \delta^{a3}$, and the resulting 
${\cal G}$ is 
the diagonal colour-space matrix 
\begin{equation}
{\cal G}^{AB}\ =\ \hbox{diag}\big(\ {\cal G}^{00}\,,\,  {\cal G}^{ab}\ \big) \ = \ 
\hbox{diag}\big(\ \frac{1}{ p^2}\,,\,  \frac{1}{ p^2-M_W^2}\,,\,
\frac{1}{ p^2-M_W^2}
\,,\, \frac{1}{p^2}\ \big)
\ \ .
\label{calGdef}
\end{equation}
The first entry corresponds to the $U(1)$ subgroup associated 
to the generator $t^0$, the last to the $a=b=3$ 
component and $M_W$ is the mass of the $W^{\pm}$ bosons, 
$M_W = \sqrt{2} |{\rm v}|$.
The one-loop perturbation theory goes on as in the last section, with the only 
modification of using as propagators the appropriate $U(N)$ generalization 
of (\ref{calGdef}). 

Next we ask whether in the spontaneously broken theory there are new IR/UV 
mixing effects compared to the unbroken case studied in the previous subsection.
To answer this question, 
we need to look only at the very high loop momentum contribution 
to the Feynman amplitudes, 
which is responsible for the interplay between ultraviolet and 
infrared divergences
\cite{Matusis}. 
In this approximation all the masses and external momenta can be ignored
which means that in the IR
we get precisely the same decoupling pattern of the $U(1)$
and no new IR/UV mixing effects.
Thus, there is no IR/UV mixing in the $SU(N)$ theory, which
means that the noncommutative $SU(N)$ behaves in the same way in the IR
as its commutative counterpart.

\section{Multi-instanton contributions to the prepotential}

In this Section we will explain why all the multi-instanton
contributions to the prepotential in the noncommutative theory
precisely agree with those computed in the
ordinary commutative theory. 

We consider the $\N=2$ supersymmetric $U(N)$ gauge theory directly
in Euclidean\footnote{For $\N>1$ there are no problems with
Euclidean formulations of supersymmetry.} 
noncommutative space. 
It will be convenient to parametrize the six independent components
of $\theta^{\mu\nu}$ in terms of the self-dual and the anti-self-dual
combinations
\EQ{
\zeta_{\sst(+)}^c \equiv \bar{\eta}^c_{\mu \nu} \theta^{\mu \nu} \ ,
 \qquad
\zeta_{\sst(-)}^c \equiv \eta^c_{\mu \nu}\theta^{\mu \nu} \ ,
\qquad c=1,2,3 \ , \label{zedefs}
}
where $\bar{\eta}^c_{\mu \nu}$ and $\eta^c_{\mu \nu}$ are the standard
self-dual and anti-self-dual 't Hooft symbols.
Note that in Euclidean space all the six components
of  $\{\zeta_{\sst(+)}^c, \zeta_{\sst(-)}^c\}$ are real and 
independent.

We now turn to the effective Lagrangian \eqref{WilsF}.
The general superfield expression on the right hand side of \eqref{WilsF}
can be expanded in component-fields and will
contain the characteristic 4-fermion and 4-anti-fermion interactions:
\SP{
\L_{\rm eff} \ &\ni \ {1 \over 32\pi i} \left\{
\partial^4_{\vhiggs} \F(\vhiggs, \zeta_{\sst(+)}, \zeta_{\sst(-)})
\,\lambda \star \lambda \star \lambda \star\lambda \ - \
\partial^4_{\vhiggsbar} \F(\vhiggs, \zeta_{\sst(+)}, \zeta_{\sst(-)})^\dagger
\,\lambdabar \star \lambdabar \star \lambdabar \star\lambdabar \right\}\\
& = \ {1 \over 32\pi i} \left\{
\partial^4_{\vhiggs} \F(\vhiggs, \zeta_{\sst(+)}, \zeta_{\sst(-)})
\,\lambda \star \lambda \star \lambda \star\lambda \ - \
\partial^4_{\vhiggsbar} \F^*(\vhiggsbar, \zeta_{\sst(+)}, \zeta_{\sst(-)})
\,\lambdabar \star \lambdabar \star \lambdabar \star\lambdabar \right\}
 \ ,
\label{fourfaf}}
where $\lambdabar(x)$ and $\lambda(x)$ are the (anti)-gauginos of the
$\N=2$ theory. The  dagger, $\F^\dagger,$ on the first line
of \eqref{fourfaf} denotes the Hermitean conjugation
which also conjugates the argument of the function,
and the asterisck, $\F^*,$ 
on the second line on \eqref{fourfaf} denotes complex conjugation
of the function without complex-conjugating the argument.\footnote{
For example, if $f(x)= 1+ix,$ then $f(x)^\dagger=1-i\bar{x},$
and $f^*(x)=1-ix,$ such that
$f(x)^\dagger = f^*(\bar{x}).$}

It is well-known \cite{Seiberg,mo2} that instantons contribute to the
4-fermion vertex
whereas anti-instantons contribute to the
4-anti-fermion vertex in the effective action:
\SP{
\L_{\rm eff} \ &\ni \ G_{\rm inst}
\,\lambda \star \lambda \star \lambda \star\lambda \ + \
G_{\rm anti\hbox{-}inst}
\,\lambdabar \star \lambdabar \star \lambdabar \star\lambdabar 
 \ ,
\label{inaincon}}
where  $G_{\rm inst}$ and 
$G_{\rm anti\hbox{-}inst},$ can be determined from the Green functions
$$ \langle \lambdabar\lambdabar\lambdabar\lambdabar \rangle_{\rm inst} \ ,
\qquad 
\langle \lambda\lambda\lambda\lambda \rangle_{\rm anti\hbox{-}inst} \ ,$$
computed in the instanton and the anti-instanton backgrounds.
The prepotential $\F$ can be now recovered in two independent ways.
The first is by relating
$\partial^4_{\vhiggs} \F(\vhiggs, \zeta_{\sst(+)}, \zeta_{\sst(-)})$
to $G_{\rm inst},$ the second, by relating  
$\partial^4_{\vhiggsbar} \F^*(\vhiggsbar, \zeta_{\sst(+)}, \zeta_{\sst(-)})$
to $G_{\rm anti\hbox{-}inst}$.

In general $G_{\rm inst}$ and $G_{\rm anti\hbox{-}inst}$ depend
on the VEVs and on the noncommutativity parameters. 
As will be explained in the next Section,
the latter dependence is very restrictive: $G_{\rm inst}$ depends
on $\zeta_{\sst (+)}$ and not on $\zeta_{\sst (-)},$
and $G_{\rm anti\hbox{-}inst}$ depends
on $\zeta_{\sst (-)}$ and not on $\zeta_{\sst (+)}.$ This is the consequence
of the fact that the (anti)-self-dual solutions in noncommutative
Yang-Mills do not depend on the noncommutativity parameter of the opposite
duality \cite{Nekrasov:1998ss, SWnc} as it is immediately realized 
by looking at the expressions for the ADHM constraints.
 From this we conclude that
$\F$ cannot depend on $\zeta_{\sst (-)},$ and $\F^*$ cannot
depend on $\zeta_{\sst (+)}.$  This means that $\F$ does not
depend neither on $\zeta_{\sst (-)},$ nor on $\zeta_{\sst (+)}.$

In fact it is easy to argue that the prepotential $\F$ 
in the noncommutative theory can be calculated
directly at $\zeta_{\sst (\pm)}=0$ leading to an
expression for $\F$ which is identical to the ordinary commutative case.
Let us set $\zeta_{\sst (+)}=0$ and keep $\zeta_{\sst (-)}\neq 0$.
The instanton contribution is then identical to the commutative theory,
as the instanton itself and the instanton measure
coincide with their commutative counterparts. The instanton
prediction for $\F$ is the same as in the commutative theory
and must match with the anti-instanton prediction for $\F$.
But the anti-instanton is truly noncommutative, as $\zeta_{\sst (-)}\neq 0$.
From this matching it follows that the general
noncommutative (anti)-instanton contribution to $\F$ can be computed
at $\zeta_{\sst (\pm)}=0$ leading to the ordinary commutative
prepotential.

In the next Section we check this general argument
against an explicit one-instanton calculation.

\section{Instanton calculations}

In this Section we introduce the  necessary tools to perform
explicit multi-instanton calculations in  noncommutative
$\N=2$ theories. We then evaluate the one-instanton contribution
to $\F$ and confirm the general argument presented in the
previous Section. 

\subsection{Multi-instanton supermultiplet}

The multi-instanton configuration in the noncommutative $\N=2$
supersymmetric $U(N)$ Yang-Mills theory on the Coulomb
branch is a (constrained) solution of the  equations
of motion of the theory in noncommutative Euclidean space-time.

Let us first consider a pure noncommutative
(nonsupersymmetric) $U(N)$ gauge theory. 
The $k$-(anti)-instanton gauge field is the
general solution of the (anti)-self-duality equations 
with instanton charge $\pm k$.
This (anti)-self-dual gauge configuration follows from the ADHM analysis
\cite{ADHM,CFGT,CWS} and can be conveniently parametrized by the 
$[N+2k]\times[2k]$ matrix of instanton collective coordinates 
$a_{\sst [N+2k]\times [2k]}$. This matrix can be written as \cite{KMS}
\begin{equation}
a_{\sst [N+2k]\times [2k]} \ = \ 
\begin{pmatrix} w_{\sst [N]\times [2k]} \\  a'_{\sst [2k]\times
[2k]}\end{pmatrix} \ = \ 
\begin{pmatrix}  w_{u \, i \dalpha}\\ 
(a'_{\beta \dalpha})^{ }_{li}\end{pmatrix}_{\phantom{q}} 
\ .
\label{canform}\end{equation}
In this Section we will closely follow notation and conventions
adopted in the Instanton Hunter's Guide \cite{KMS} to which the reader
is referred for more detail on the instanton calculus. 
In particular we use the following index assignments:
\begin{align}\hbox{Instanton number indices\ }[k]:\qquad&1\le i,j,l\cdots\le
k&\notag \\
\hbox{U(N) Color indices\ }[N]:\qquad&1\le u,v\cdots\le N&\notag \\
\hbox{ADHM indices\ }[N+2k]:\qquad&1\le \lambda,\mu\cdots\le N+2k&\notag \\
\hbox{Quaternionic (Weyl) indices\ }[2]:\qquad&\alpha,\beta,\dalpha,
\dbeta\cdots=1,2&\notag 
\end{align}

Importantly, not all the components of the ADHM matrix $a$ are independent.
Much of this redundancy
can be eliminated by noting that the ADHM construction 
of the gauge field is
unaffected by $x$-independent $U(k)$ transformations of the form
\begin{equation}w_{u  i\dalpha}\, \ \to \ w_{u  j\dalpha}\, R_{ji}
\ , \quad 
(a'_{\alpha\dalpha})_{ij} \ \to \ R^\dagger_{il} \ 
(a'_{\alpha \dalpha})_{lp}\ R_{pj} \ , \quad R_{ij} \in U(k)
\ .
\label{restw}\end{equation}
Finally, and most importantly,
the self-duality equations require that 
\begin{subequations}
\begin{align}
\trtwo\,\tau^c \bar{a} a \ &= \ 0 \label{fconea}\\
(a^{\prime}_n)^\dagger  \ &= \ a^{\prime}_n\ .
\label{fconeb}\end{align}
\end{subequations}
In Eq.~\eqref{fconea} we have contracted $\bar{a}^\bD a_\aD$ with the three Pauli matrices
$({\tau^c})_{\ \dbeta}^{\dalpha}$, while in Eq.~\eqref{fconeb} we have decomposed
 $(a'_{\alpha\dalpha})_{li}$ and $(\bar{a}^{\prime\dalpha\alpha})_{il}$
in the usual quaternionic basis of spin matrices: 
\begin{equation}(a'_{\alpha \dalpha})^{}_{li} \ = \ 
(a'_\mu)^{}_{li} \ \sigma^\mu_{ \alpha\dalpha} \ , \quad
(\bar{a}^{\prime\dalpha \alpha})^{}_{il} \ = \
(a'_\mu)^{}_{il} \ \sigmabar^{\mu \, \dalpha \alpha}\ .
\label{dec}\end{equation}
Equation \eqref{fconea} is the famous non-linear matrix equation
which is frequently referred to as the ADHM constraint.
We can count the 
independent bosonic collective coordinates of the ADHM 
multi-instanton solution. A general complex matrix 
$a_{\sst [N+2k]\times [2k]}$ has $4k(N+2k)$ real degrees of freedom,
$n_{\rm b}$.
The two ADHM conditions \eqref{fconea} and \eqref{fconeb} impose
$3k^2$ and $4k^2$ real constraints, respectively, while modding out
by the residual $U(k)$ symmetry removes another
 $k^2$ degrees of freedom. In total we therefore have
\begin{equation}
n_{\rm b}\, \equiv\, 4k(N+2k)-3k^2-4k^2- k^2 \, = \, 4kN 
\label{dofb}\end{equation}
real degrees of freedom, precisely as required.

We now can return to the gauge theory on noncommutative space.
The noncommutative multi-instanton configuration can be obtained by a 
straightforward modification of the ordinary ADHM construction.
Nekrasov and Schwarz \cite{Nekrasov:1998ss} showed that in noncommutative space
the (anti)-instanton
ADHM constraint \eqref{fconea} is shifted by
the (anti)-self-dual component
of $\theta^{\mu\nu}$
\AL{
{\rm instanton} \ : \qquad
&\trtwo ( \tau^c \bar{a} a)_{ij} - \zeta_{\sst(+)}^{c} \delta_{ij}\ = \ 0 \ , \qquad
\zeta_{\sst(+)}^c \equiv \bar{\eta}^c_{\mu \nu} \theta^{\mu \nu} \label{msd}\\
{\rm anti\hbox{-}instanton} \ : \qquad
&\trtwo ( \tau^c \bar{a} a)_{ij} - \zeta_{\sst(-)}^c \delta_{ij}\ = \ 0 
 \ , \qquad
\zeta_{\sst(-)}^c \equiv \eta^c_{\mu \nu}\theta^{\mu \nu} \label{masd}
}
In $\N=2$ supersymmetric gauge theories the gauge field is accompanied by
two gauginos, $\lambda^A$, $A=1,2$
and a complex Higgs field, all in the adjoint representation of
$U(N)$. The corresponding instanton component fields were determined
in \cite{KMS}. The instanton components of gauginos are traditionally
referred to as the adjoint fermion zero modes. They have an associated
set of Grassmann collective coordinates which can be arranged into the
$[N+2k]\times[k]$ matrices $\M^A$ and $\bar{\M}^A$ as in \cite{KMS}
\begin{equation}
\M^A_{\sst [N+2k]\times[k]}  \ = \
\begin{pmatrix} \mu^A_{u i} \\  (\M^{\prime A}_\beta)_{li}\end{pmatrix}   \ ,\qquad
\Mbar^{A}_{\sst [k] \times [N+2k]} \ = \ 
\left( \mubar^A_{i u} \ ,\ (\Mbar^{\prime\beta A})_{il} \right)\ .
\label{mrep}\end{equation}
Dirac equations for the fermion zero modes in the ADHM background
require the matrices $\M^{A}$
and $\Mbar^{A}$ to satisfy the so-called fermionic ADHM constraints:
\begin{subequations}
\begin{align}
\Mbar^{ A}_{i} \, a_{ j\dalpha}
 \ &= \ - \bar{a}_{i\dalpha} \, \M^A_{ j}
\ ,\label{zmcona}\\
\Mbar^{\prime A}_\alpha\ &= \
\M^{\prime A}_\alpha
\ .  
\label{zmconb}\end{align}
\end{subequations}
Equation \eqref{zmconb} allows us to eliminate $\Mbar^{\prime A}$ in favour of
$\M^{\prime A}$.
Counting the number of fermionic degrees of freedom for the first gaugino, 
$\lambda^1$, one finds $2k(N+2k)$ 
real Grassmann parameters in  $\M^1$ and $\Mbar^1$, subject to $2k^2$ 
constraints from 
each of Eqs. \eqref{zmcona}, \eqref{zmconb} for a net of $2Nk$ gaugino
zero modes as required. The same counting applies to the second fermion
flavour, $\lambda^2$, with the total net effect of
\EQ{ n_{\rm f} \ \equiv \ 4Nk \ . \label{fdof}}
For future reference we note here, following \cite{KMS}, that the
instanton supermultiplet also contains
the anti-gaugino components, $\bar{\lambda}^A$, which, however, do not
lead to new Grassmann collective coordinates. 

Finally, the adjoint Higgs field configuration is constructed in \cite{KMS}
in terms of the auxiliary 
$k\times k$ anti-Hermitian matrix $\Atot$ which is defined  as the
solution to the inhomogeneous linear  equation
\EQ{
\bigL\cdot\Atot \ =\ \Lambdatot \ ,
\label{thirtysomething}}
where $\Lambdatot$ is the $k\times k$ anti-Hermitian matrix
\def\Lambdabar{\bar\Lambda}
\EQ{
{\Lambdatot}_{ij}\ =\ \wbar^\dalpha_{i u} \ \homoa_{uv} \ 
w_{v j\dalpha} \ + \ 
{1\over2\sqrtwo}\,
\big(\,\Mbar^1\M^2 -\Mbar^2\M^1 \,\big)_{ij}
\ , 
\label{Lambdabardef}
}
and the $N\times N$ matrix $\homoa$ is just $i$ times the VEV matrix,
\EQ{
{\homoa}_{uv}
\ =\ i \ {\rm diag}(\vhiggs_{\sst 1}, \ldots, \vhiggs_{\sst N})
 \ . \label{vevbcagain}}
$\bigL$ is a linear operator that maps the space of $k\times k$ 
 scalar-valued anti-Hermitian matrices onto itself. Explicitly,
if $\Omega$ is such a matrix, then $\bigL$ is defined as 
\EQ{
\bigL\cdot\Omega\ =\ 
\hf\{\,\Omega\,,\,W\,\}\,-\,\hf\trtwo\big(
[\,\bar{a}'\,,\,\Omega\,]a'-\bar{a}'[\,a'\,,\,\Omega\,]\big)
\label{bigLreally}}
where 
$W$ is the Hermitian  $k\times k$ matrix
$W_{ij}=\wbar^\dalpha_{i u} w_{u j\dalpha}$.
Note that the matrix  $\A$ is completely determined by the 
inhomogeneous equation
\eqref{thirtysomething}, as a result, there are no (unconstrained) collective
coordinates associated with the Higgs. However, we can still interpret
the $\A$ as the (constrained)
collective coordinates for the Higgs field, subject to the
`ADHM Higgs constraint'
\eqref{thirtysomething}. 

The $k$-instanton action in the $\N=2$ supersymmetric $U(N)$ gauge
theory was calculated in \cite{KMS}. It reads:
\SP{
S^{(k)} \ &= \ 
{8 k \pi^2 \over g^2} \ + \ 
8\pi^2 \ \wbar^\dalpha_{iu} \barhomoa_{uu} \homoa_{uu}
 w_{ui\dalpha} \ - \ 
8\pi^2 \ \wbar^\dalpha_{i u} \barhomoa_{uu} w_{u j\dalpha} (\Atot)_{ji} 
\\
&+ \ 
2\sqrtwo \pi^2 \big( \mubar^1_{iu} \barhomoa_{uu} \mu^2_{ui}
\ - \ \mubar^2_{iu} \barhomoa_{uu} \mu^1_{ui}  \big)  \ ,
\label{siact}}
where 
${\barhomoa}_{uv}
\ =\ -i \ {\rm diag}(\bar{\vhiggs}_{\sst 1}, \ldots,\bar{\vhiggs}_{\sst N})$

Similar considerations apply to the
anti-instanton supermultiplet, which will have 
fermion and  anti-fermion components interchanged,
and $\homoa$ exchanged with $\barhomoa$.
The anti-instanton action is then
\SP{
S^{(-k)} \ &= \ 
{8 k \pi^2 \over g^2} \ + \ 
8\pi^2 \ \wbar^\dalpha_{iu} \barhomoa_{uu} \homoa_{uu}
 w_{ui\dalpha} \ - \ 
8\pi^2 \ \wbar^\dalpha_{i u} \barhomoa_{uu} w_{u j\dalpha} (\Atot)_{ji} 
\\
&+ \ 
2\sqrtwo \pi^2 \big( \mubar^1_{iu} \homoa_{uu} \mu^2_{ui}
\ - \ \mubar^2_{iu} \homoa_{uu} \mu^1_{ui}  \big)  \ ,
\label{saiact}}
where the Grassmann collective coordinates $\mu$, $\mubar$ and
$\M'$ correspond to the antifermion zero modes.

\subsection{Multi-(anti)-instanton measure}

The collective-coordinate $(\pm k)$-instanton integration measure 
$d\mu^{(\pm k)}$ for a
noncommutative $\N=2$ supersymmetric $U(N)$ Yang-Mills on the Coulomb
branch is easily obtained from the measure in the ordinary commutative
theory, derived in \cite{KMS}. 

It can be argued in parallel with Seiberg and Witten \cite{SWnc},
that the only effect of 
noncommutativity on the (anti)-instanton measure 
and the action
is the shift of the gauge-field-ADHM constraint \eqref{fconea}
as described by \eqref{msd} and \eqref{masd}. 
In particular, in the noncommutative ${\cal N}=2$ supersymmetric
theory, the measure has the following form cf. \cite{KMS}:
\begin{equation}\begin{split}
\int d\mu^{(\pm k)} \ \exp[-S^{(\pm k)}]
\ &= \ {M_{\sst PV}^{2Nk}(C'_1)^k \over{\rm Vol}\,U(k)}
\int d^{4k^2} a' \
d^{2kN} \wbar \ d^{2kN} w  
\prod_{A=1,2}
\ d^{2k^2} \M^{\prime A}  
\ d^{kN} \mubar^A \ d^{kN} \mu^A\\
&\times 
\ d^{k^2} \Atot \
\prod_{c=1,2,3}
\delta^{(k^2)}\big(\tfrac12(\trtwo\, \tau^c \bar{a} a -\zeta_{\sst (\pm)} )\big)
\prod_{A=1,2}
\delta^{(2k^2)}(\Mbar^A a + \bar{a} \M^A
)\\ &\times
\delta^{(k^2)}(\bigL\cdot\Atot - \Lambdatot)\ \exp[-S^{(\pm k)}] \ .
\label{dmudef}\end{split}\end{equation}
Here the integrals on the right hand side of \eqref{dmudef}
are over all the collective coordinates
of the instanton supermultiplet. The ADHM constraints for gauge field \eqref{msd} and \eqref{masd},
fermions \eqref{zmcona}, and 
the Higgs \eqref{thirtysomething},
are explicitly implemented via the delta-functions.

We stress that the noncommutativity parameters $\zeta_{\sst (\pm)}$
appears in \eqref{dmudef} only via the expression
$\delta^{(k^2)}\big(\tfrac12(\trtwo\, 
\tau^c \bar{a} a -\zeta_{\sst (\pm)} )\big).$ All the other factors in 
the instanton partition function \eqref{dmudef} (including the other constraints
and expression for the instanton action \eqref{siact}) are unchanged.
One way to understand this is to appeal to the $k$-D-instanton 
partition function in the presence of $N$ D3-branes
in type IIB string theory, which was derived in Section IV.2 of \cite{mo3}.
In the $\alpha'\to 0$ limit this D-instanton
partition function reduces to the Yang-Mills-instanton partition function
in $\N=4$ supersymmetric Yang-Mills on the world-volume of D3-branes. 
As explained in \cite{SWnc}, the noncommutativity on the world-volume of D3-branes is introduced by turning on a background $B_{\mu\nu}$ field.
In the $k$-D-instanton matrix theory this corresponds to turning on
the Fayet--Iliopolos (FI) couplings in the $U(1)$ subgroup of the $U(k)$
gauge theory, which leads precisely to the modification of the
gauge-field-ADHM constraints \eqref{msd} and \eqref{masd}.
Calculations of instanton partition functions in $\N=4$ with
and without noncommutativity were performed in \cite{Partition}.
Finally,
$\N=4$ supersymmetry can be then softly broken by mass terms to
$\N=2$, leading precisely \cite{mo3} to \eqref{dmudef}.

The factor of $C'_1$ on the right hand side of \eqref{dmudef}
is a numerical constant associated with the normalization
of the 1-instanton measure, and
$M_{\sst PV}^{2Nk}$ corresponds to the Pauli-Villars regulator to the power
$n_{\rm b}-\tfrac12 n_{\rm f}=2Nk$ which arises from the ratio
of the UV-regularized bosonic and fermionic fluctuation determinants.
Combined with $\exp[-8k\pi^2/g^2(M_{\sst PV})]$ from the instanton action
it gives rise to the renormalization group invariant 
scale $\Lambda_{\sst PV}$
of the $\N=2$ noncommutative $U(N)$ theory,
\EQ{
M_{\sst PV}^{2Nk} \ \exp\left[-{8k\pi^2\over g^2(M_{\sst PV})}\right] \ = \
\Lambda_{\sst PV}^{2Nk} \ \equiv \ \Lambda_{\sst PV}^{b_0 k} \ .
\label{dscale}
}

\subsection{Explicit expression for the $\N=2$ prepotential}

The general expression for the $k$-instanton 
contribution to the prepotential \eqref{multiexp} was derived in
\cite{dkmmod, mo2, KMS}
\EQ{
{\cal F}^{(k)}(\vhiggs)
\ =\ 8\pi i\int d \tilde{\mu}^{(k)}\,\exp[-S^{(k)}]
\label{Fnfinal}}
Here $d \tilde{\mu}^{(k)}$ is the ``reduced measure''
which is obtained from the $\N=2$ measure,
$d{\mu}^{(k)}$, as follows:
\EQ{
\int d{\mu}^{(k)}\ =\ \int d^4x_0\,d^2\xi_1\,d^2\xi_2\,
\int d \tilde{\mu}^{(k)} \ ,
\label{mufactor}}
where $(x_0,\xi_1,\xi_2)$ gives the global position of the 
multi-instanton in $\N=2$ superspace. Explicitly, the instanton
center $x^\mu_0$ and the supersymmetric fermion zero modes  $\xi_1,$
$\xi_2$ are the linear combinations proportional to the ``trace'' components
of the $k\times k$ matrices $a',$ $\M^{\prime 1}$ and $\M^{\prime 2}$,
respectively:
\EQ{
x_0\ =\ {1\over k}\Tr_k\,a'\ ,\quad
\xi_1\ =\ {1\over4k}\Tr_k\M^{\prime 1}\ ,\quad
\xi_2\ =\ {1\over4k}\Tr_k\M^{\prime 2}\ .
\label{traceparts}}
Note that these $\N=2$ superspace modes do not enter into the
$\delta$-function constraints
and so do indeed factor out in this simple way. Furthermore, the four
exact supersymmetric fermion zero modes $\xi_{1\alpha}$ and 
$\xi_{2\alpha}$ are the only fermionic modes that are not lifted by
(\ie\ do not appear in) the action \eqref{siact}.

Given these expressions for the prepotential, one also knows the
all-instanton-orders expansion of the coordinate on the vacuum moduli
space of the theory
$u_2=\langle\Tr\,A^2\rangle$, since on general grounds 
\EQ{
u_2(\vhiggs){{}\atop\Big|}_{k\hbox{-}{\rm inst}}\ 
=\ 2i\pi k\cdot
\F^{(k)}(\vhiggs) \ .
\label{matrelis}}
This relation was originally derived by Matone
\cite{Matone}, further studied at the $2$-instanton level by \cite{FT},
and the all-instanton-orders  proof 
of it was presented in \cite{dkmmod}.
The above collective coordinate integral expression for $\F^{(k)}$
constitutes
a closed series solution, in quadratures, of the low-energy dynamics of the
Coulomb branches of the
$\N=2$ models. It is noteworthy that this solution
is obtained purely from the instanton physics.

\subsection{One-instanton contribution to the prepotential}

In this section we will explicitly evaluate the 1-instanton contribution
to the prepotential, $\F^{(1)}$ in the noncommutative theory
with non-vanishing $\zeta_{\sst +}$. We will see that this
expression will be exactly the same as in the commutative theory,
in agreement with the general argument of the previous section.

Our starting point is the
integral expression \eqref{Fnfinal} in terms of the reduced noncommutative 
instanton measure. Our analysis will follow
closely the commutative instanton calculation presented in Section 8
of \cite{KMS}. To evaluate the integral on the right hand side of
\eqref{Fnfinal}, it is helpful to exponentiate the various
$\delta$-functions by means of Lagrange multipliers, and to interchange
the resulting order of integration. In other words, one integrates
out the ADHM supermultiplet $\{a,\M^1,\M^2,\A\}$ first, 
and only then performs
the integration over the Lagrange multipliers.

In the 1-instanton case the spin-1 and spin-$1/2$
ADHM constraints involve only the top-row elements of matrices
$a$ and $\M$. They can be
exponentiated in a simple way
\def\bp{{\bf p}}
\EQ{
\prod_{c=1,2,3}
\delta\big(\hf(\tau^c)^\dalpha_{\ \dbeta}\,\wbar_u^\dbeta
w^{}_{u\dalpha} -\hf \zeta^c_{\sst (+)}\big)\ =\ 
{1\over\pi^3}\,\int d^3\bp\,
\exp(ip^c(\tau^c\wbar_u w_u -\zeta^c_{\sst (+)})\ ,
\label{spinone}}
and
\AL{
\prod_{\dalpha=1,2}
\delta\big(\mubar^1_uw_{u\dalpha}+\wbar_{u\dalpha}\mu^1_u\big)
\ &=\ 2\int d^2\xi\,\exp\big(\xi^\dalpha(\mubar^1_uw_{u\dalpha}
+\wbar_{u\dalpha}\mu^1_u)\big)
\label{spinhfa}
\\
\prod_{\dalpha=1,2}
\delta\big(\mubar^2_uw_{u\dalpha}+\wbar_{u\dalpha}\mu^2_u\big)
\ &=\ 2\int d^2\eta\,\exp\big(\eta^\dalpha(\mubar^2_uw_{u\dalpha}
+\wbar_{u\dalpha}\mu^2_u)\big)
\label{spinhfb}
}
In this way we introduce the triplet of bosonic Lagrange multipliers
$p^c$, as well as the Grassmann spinor Lagrange multipliers
$\xi^\dalpha$ and $\eta^\dalpha.$ 
The exponentiation of the spin-0 constraint is best accomplished 
involving a term in the action \eqref{siact}
$ 8\pi^2 \ \wbar_{u} \barhomoa_{uu} w_{u} \Atot 
\equiv 8\pi^2\Lambdabar\Atot$  
as follows:
\def\Re{{\rm Re\,}}
\def\zbar{{\bar z}}
\def\Im{{\rm Im\,}}
\SP{
&\int 
d\Atot\,\delta(\bigL\cdot\Atot-\Lambdatot)\,
\exp(8\pi^2\Lambdabar\Atot)\ =\ {1\over\det\bigL}\,
\exp(8\pi^2\Lambdabar\cdot\bigL^{-1}\cdot\Lambda)
\\
&=\ 8\pi\int d(\Re z)d(\Im z)\,\exp\big(-8\pi^2(\zbar\,\bigL\, z-
\Lambdabar z-\zbar\Lambdatot)\big)
\label{spinzero}}
The  second equality  follows from the general Gaussian identity
\EQ{\int\prod_i
d(\Re z_i)d(\Im z_i)\,\exp\big(-\zbar_iK_{ij}z_j+\bar y_iz_i+\zbar_iy_i\big)
\ =\ {1\over\det(K/\pi)}\,\exp(\bar y_iK^{-1}_{ij}y_j)
\label{gaussid}}
which can be used to exponentiate the spin-0 constraint in an elegant
way for arbitrary instanton number $k$. 
The advantage of the rewrite \eqref{spinzero} is that $\bigL$ is easier to
manipulate in the exponent than $\bigL^{-1}$ (which appears implicitly 
in the definition of $\Atot$).
In the present case, with $k=1,$
the operator $\bigL$ collapses to a $1\times1$ $c$-number matrix:
\EQ{\bigL\ =\ \det\bigL\ =\ \wbar_u^\dalpha w^{}_{u\dalpha} \equiv
2 \rho^2\ ,
\label{bigLcollapse}}
where $\rho$ is the instanton size.
Likewise $\Lambdabar$ and $\Lambdatot$ are given by
\EQ{
\Lambdabar=-i\vhiggsbar_u\wbar_u^\dalpha w^{}_{u\dalpha}\ ,
\qquad \Lambdatot=i{\rm v}_u\wbar_u^\dalpha w^{}_{u\dalpha}-{1\over2\sqrtwo}
(\mubar^2_u\mu^1_u-\mubar^1_u\mu^2_u)\ .
\label{alsocollapse}}

\def\alphabar{\bar\alpha}
Now we can perform the Grassmann
integrations over  $\{\mu^1_u,\mu^2_u,\mubar^1_u,\mubar^2_u\}$.
Consider the combined exponent formed from 
Eqs.~\eqref{integraldef}-\eqref{spinzero} and the remaining terms
in the 1-instanton action, 
\EQ{
\exp[-S^{(1)}] \ni \exp \big[-8\pi^2|{\rm v}_u|^2\wbar_u^\dalpha w^{}_{u\dalpha}+2\sqrtwo\pi^2i
(\mubar^1_u\bar{\rm v}_u\mu^2_u-\mubar^2_u\bar{\rm v}_u\mu^1_u)\big] \ .
\label{oneiactn}}
To eliminate the linear terms in Grassmann variables in the
combined exponent, we first perform the linear shifts
\SP{
\mu^1_u&\rightarrow\mu^1_u+{i\eta^\dalpha w_{u\dalpha}\over
2\sqrtwo\pi^2\alphabar_u},\quad
\mubar^1_u\rightarrow\mubar^1_u+{i\eta^\dalpha \wbar_{u\dalpha}\over
2\sqrtwo\pi^2\alphabar_u},\quad\cr
\mu^2_u&\rightarrow\mu^2_u-{i\xi^\dalpha w_{u\dalpha}\over
2\sqrtwo\pi^2\alphabar_u},\quad
\mubar^2_u\rightarrow\mubar^2_u-{i\xi^\dalpha \wbar_{u\dalpha}\over
2\sqrtwo\pi^2\alphabar_u}
\label{linshift}}
and then perform straightforward integrations over the remaining quadratic
terms. By inspection, the Grassmann
integrations simply bring down a factor of 
\EQ{
\prod_{u=1}^N(2\sqrtwo\pi^2i\alphabar_u)^2
\label{prodown}}
In \eqref{linshift}-\eqref{prodown}, we have
 defined $\alpha_u$ and $\alphabar_u$ as the naturally appearing
linear combinations
\EQ{
\alpha_u\ =\ {\rm v}_u+iz\ ,\qquad\alphabar_u\ =\ \vhiggsbar_u-i\zbar
\ .
\label{lincomb}}
Next, the $\{w_u,\wbar_u\}$ integrations are accomplished, using the
identity
\EQ{
\int d^2w_ud^2\wbar_u\,\exp\big(-A^0\wbar_u^\dalpha
w^{}_{u\dalpha}+i\sum_{c=1,2,3}A^c(\tau^c)^\dalpha_{\ \dbeta}\,\wbar_u^\dbeta
w^{}_{u\dalpha}\big)\ =\ {-4\pi^2\over (A^0)^2+\sum(A^c)^2}\ .
\label{wintid}}

In this way, all the original
ADHM variables $\{a,\M,\N,\Atot\}$ are eliminated
from the integral. One is left with an integral
over Lagrange multipliers only
\EQ{
\Foneinst \ = i
{C_1' \over2\pi^2}\,\int d^3\bp\, d^2\xi\, d^2\eta
\,d({\rm Re}\, z)d({\rm Im}\, z)\,
{\cal B}e^{-i\bp\cdot\bzeta}\ ,
\label{integraldef}
}
where
\EQ{
{\cal B}=\prod_{u=1}^N\ {(2\sqrtwo\pi^2 i\alphabar_u)^2(-4\pi^2)\over
\big(8\pi^2|\alpha_u|^2\big)^2+\sum_{c=1,2,3}\,(p^c+\Xi^c_u)^2}
}
and $\Xi^c_u$ is the fermion bilinear
\EQ{
\Xi^c_u={1\over4\sqrtwo\pi^2\alphabar_u}
\big(\xi^{}_\dalpha(\tau^c)^\dalpha_{\ \dbeta}\,\eta^\dbeta-
\eta^{}_\dalpha(\tau^c)^\dalpha_{\ \dbeta}\,\xi^\dbeta\big)\ .
}
The expression \eqref{integraldef} is analogous to Eq.~(8.13) of
\cite{KMS}, except here we are taking the pure gauge case,
$N_F=0$, and there is an additional phase $e^{-i\bp\cdot\bzeta}$
that arises from the noncommutativity parameter, 
$\bzeta \equiv \{\zeta^1_{\sst(+)},\zeta^2_{\sst(+)},\zeta^3_{\sst(+)}\}.$

The $\{\xi,\eta\}$ Grassmann integrations in \eqref{integraldef}
must be saturated with two insertions of $\Xi$:
\EQ{
\int d^2\xi d^2\eta\,\Xi^b_u\,\Xi^c_v\ =\
{\delta^{bc}\over16\pi^4\alphabar_u\alphabar_v}\ .}
Extracting these quadratic powers of $\Xi$ from $\calB$ can be done
quite elegantly, thanks to the algebraic identity
\SP{\int d^2\xi d^2\eta\,\calB\ &=\
\sum_{b,c=1}^3\,\sum_{u,v=1}^N{\delta^{bc}\over16\pi^4\alphabar_u\alphabar_v}
\cdot{1\over2}{\partial^2\over\partial\Xi^b_u\,\partial\Xi^c_v}\,\calB
{\Big|}_{\Xi=0}\\
&=\ {1\over32\pi^4|\bp|^2}\,\Big(\sum_{u=1}^N{\partial\over\partial
\vhiggsbar_u}\Big)^2\,\calB{\Big|}_{\Xi=0}\ .\label{niceid}}
Pulling the VEV derivatives outside the integral, one therefore finds:
\EQ{\Foneinst\ =\
{iC_1'\over2\pi^2}\cdot{1\over32\pi^4}
\Big(\sum_{u=1}^N{\partial\over\partial\vhiggsbar_u}\Big)^2
\,\int d({\rm Re}\, z)d({\rm Im}\, z)\,\Gamma\ .\label{integralb}}
Here
\EQ{\Gamma\ =\ \int d^3\bp\,{e^{-i\bp\cdot\bzeta}\over|\bp|^2}\,
\prod_{u=1}^N\ {(2\sqrtwo\pi^2 i\alphabar_u)^2(-4\pi^2)\over
\big(8\pi^2|\alpha_u|^2\big)^2+|\bp|^2}\ .
}
the angular integrals of $\bp$ are easily done,
\EQ{
\int d(\cos\theta)\,d\phi\,e^{-i\bp\cdot\bzeta}
=\frac{4\pi\sin(\zeta p)}{\zeta p}
\ ,
}
where $p\equiv |\bp|$ and $\zeta\equiv|\bzeta|$. 
The integral over $p$ can now be performed as a
standard contour integration, extended to run from $-\infty$ to $\infty$:
\EQ{
\Gamma=
16\pi^4
\Big(\frac{\pi^2}2\Big)^N\frac1{8\pi^2\zeta}
\sum_{u=1}^N
\frac1{\alpha_u^2}
\Big(\prod_{v\neq u}
{1\over \alpha_v^2}
-e^{-8\pi^2|\alpha_u|^2
\zeta}\prod_{v\neq u}
{\alphabar^2_v\over|\alpha_v|^4-|\alpha_u|^4}\Big)
\ .
\label{Gammadef}}

In this fashion, the original expression \eqref{integraldef} has collapsed
to a 2-dimensional integral over the $xy$ plane (with $x={\rm Re} z$ and
$y={\rm Im} z$ henceforth). The remaining integral may be calculated along
the lines of \cite{KMS} and so we follow that approach almost verbatim.
Notice that the only
dependence on $\vhiggsbar_u$ in the integrand is through the variables
$\alphabar_u=\vhiggsbar_u-i\zbar$. 
Therefore, it is tempting---but incorrect---to
pull the $\vhiggsbar_u$ derivatives back inside the integrand, and to
make the naive replacement
\EQ{
\sum_{u=1}^N{\partial\over\partial\vhiggsbar_u}\ \rightarrow\
i\,{\partial\over\partial\zbar}\ ,\qquad
\Big(\sum_{u=1}^N{\partial\over\partial\vhiggsbar_u}\Big)^2\ \rightarrow\
-\Big({\partial\over\partial\zbar}\Big)^2\ .
\label{replace}
}
The error here is due to the fact that the two sides of Eq.~\eqref{replace} can
differ by $\delta$-function contributions which arise at the locations
of poles in the $z$ variable. As a simple example, whereas obviously
$\big(\sum\,\partial/\partial\vhiggsbar\,\big)\,z^{-1}=0,$ one also has,
in contrast,\footnote{The normalization factor on the right-hand side
of Eq.~\eqref{exampolea} is easily fixed by integrating both sides
against $\exp(-\lambda z\zbar)$.}
\AL{
{\partial\over\partial\zbar}\,{1\over z}\ &=\ \pi\, \delta(x)\delta(y)
\ ,\label{exampolea}\\
\Big({\partial\over\partial\zbar}\Big)^2\,{1\over z}\ &=\
\pi\,{\partial\over\partial\zbar}\, \delta(x)\delta(y)\ =\
{\pi\over2}\,\big(
\delta'(x)\delta(y)+i\delta(x)\delta'(y)\big)\ .\label{exampoleb}
}
The lesson is that one can legitimately
trade $\vhiggsbar_u$ differentiation for $\zbar$
differentiation as per Eq.~\eqref{replace}---but only if one explicitly
subtracts off the
extraneous $\delta$-function pieces that are generated at
the locations of the poles in $z$. Accordingly, we can split up
$\Foneinst$ into two parts,
\def\calF{{\cal F}}
\EQ{
\Foneinst\ =\ \calF_\delta\ +\ \calF_\partial\ ,
\label{Fsplit}
}
where $\calF_\delta$
is the contribution of these $\delta$-function corrections,
while $\calF_\partial$ is a boundary term arising from judicious use
of Stokes' theorem applied to $\partial^2/\partial\zbar^2.$ Let us evaluate
each of these parts, in turn:

\def\bigR{{\rm I}\!{\rm R}}

As stated, to calculate $\calF_\delta,$ one converts $(\sum\partial/\partial
\vhiggsbar_u)^2$ into $-\partial^2/\partial\zbar^2$ as per Eq.~\eqref{replace},
then subtracts off the spurious $\delta$-function contributions that
correspond to the poles in $z$ of the expression $\Gamma$ given in
Eq.~\eqref{Gammadef}. The relevant poles lie at the $N$ distinct values
\EQ{
0\ =\ \alpha_u\ = {\rm v}_u+iz\ =\ ({\rm Re} \,{\rm v}_u-y)+i({\rm Im} 
\,{\rm v}_u+x)\
.\label{poledef}
}
There also appear to be poles in $\Gamma$ when $|\alpha_v|^2=
\pm |\alpha_u|^2$ but these are irrelevant: the poles at
$|\alpha_v|^2=-|\alpha_u|^2$ lie away from the real domain of
integration  $(x,y)\,\in\,\bigR^2,$ whereas the poles at
$|\alpha_v|^2=+|\alpha_u|^2$ have residues that cancel pairwise
among the terms in Eq.~\eqref{Gammadef} (these pairs correspond to
interchanging the indices $u$ and $v$). In the vicinity of the 
the singularity \eqref{poledef}, we have
\EQ{
\frac1{8\pi^2\zeta\alpha_u^2}
\Big(\prod_{v\neq u}
{1\over \alpha_v^2}
-e^{-8\pi^2|\alpha_u|^2
\zeta}\prod_{v\neq u}
{\alphabar^2_v\over|\alpha_v|^4-|\alpha_u|^4}\Big)
\thicksim \frac{\bar\alpha_u}{\alpha_u}\prod_{v\neq u}\,
{\alphabar^2_v\over|\alpha_v|^4-|\alpha_u|^4}+\cdots\ ,
}
which is identical to the behaviour in the case when $\zeta=0$. In
other words this means that $\calF_\delta$ is identical to the
expression derived in \cite{KMS} for the $\zeta=0$ case:
\EQ{\calF_\delta\ =\ -{iC_1'\pi^{2N-1}\over2^{N+2}}
\,\sum_{u=1}^N\,\prod_{v\neq u}\,{1\over({\rm v}_v-{\rm v}_u)^2}\,
\ .\label{Fdelfinal}
}

\def\calD{{\cal D}}
Next we consider the boundary term $\calF_\partial$ implied by the
naive replacement \eqref{replace}. It is useful to switch to polar coordinates,
$(x,y)\rightarrow(r,\theta),$ in terms of which
\EQ{
{\partial^2\over\partial\zbar^2}\ =\
{1\over r}\,{\partial\over\partial r}\circ\calD_r\ +\
{\partial\over\partial\theta}\circ\calD_\theta
\label{zbarderiv}
}
where
\EQ{\calD_r\ =\ \fourth e^{2i\theta}\,\big(\,
2+r\,{\partial\over\partial r}\,\big)\ ,\qquad
\calD_\theta\ =\ {i\over4r^2}\,e^{2i\theta}\,\big(\,
1+2r{\partial\over\partial r}
+i\,{\partial\over\partial\theta}\,\big)\ .
\label{calDdef}
}
Since the integrand in Eq.~\eqref{integralb} is a single-valued function
of $\theta,$ the $(\partial/\partial\theta)\,
\calD_\theta$ term can be neglected.
Stokes' theorem then equates the 2-dimensional integral \eqref{integralb}
to the angularly integrated action of $\calD_r$ evaluated on the circle
of infinitely large radius:
\SP{
\calF_\partial\ &=\
-{iC_1'\over2\pi^2}\cdot{1\over32\pi^4}\cdot 8\pi^6\,
\lim_{r\rightarrow\infty}\,\quarter\big(\,2+r\,{\partial\over\partial r}\,\big)
\\
&\times\ \int_0^\infty d\theta\,e^{2i\theta}\,
\Big[\,\sum_{u=1}^N\,
\frac1{8\pi^2\zeta\alpha_u^2}
\Big(\prod_{v\neq u}
{1\over \alpha_v^2}
-e^{-8\pi^2|\alpha_u|^2
\zeta}\prod_{v\neq u}
{\alphabar^2_v\over|\alpha_v|^4-|\alpha_u|^4}\Big)
\,\Big]
\ ,
\label{Fddef}
}
where $\alpha_u={\rm v}_u+ire^{i\theta}$ and
$\alphabar_u=\vhiggsbar_u-ire^{-i\theta}$. But for large $r$,
\EQ{
\frac1{8\pi^2\zeta\alpha_u^2}
\Big(\prod_{v\neq u}
{1\over \alpha_v^2}
-e^{-8\pi^2|\alpha_u|^2
\zeta}\prod_{v\neq u}
{\alphabar^2_v\over|\alpha_v|^4-|\alpha_u|^4}\Big)
\thicksim r^{-2}
}
and therefore $\calF_\partial$ vanishes. This is identical to the
value of $\calF_\partial$ in the case when $\zeta=0$ \cite{KMS}.

So finally we have proved
\EQ{
\Foneinst\ \equiv\ \calF_\delta\ =\
-{iC_1'\pi^{2N-1}\over2^{N+2}}
\,\sum_{u=1}^N\,\prod_{v\neq u}\,{1\over({\rm v}_v-{\rm v}_u)^2}\,
}
and in particular 
\EQ{
\Foneinst(\vhiggs_u,\bzeta)=\Foneinst(\vhiggs_u,\bzeta=0)\ .
}

The fact that derivatives of the prepotential are independent of
$\bzeta$ is a strong constraint on the multi-instanton
contributions. In fact it suggests
\EQ{
{\cal F}_k(\vhiggs_u,\bzeta)={\cal F}_k(\vhiggs_u,0)+{\cal S}_k(\zeta)\ ,
\label{hhh}
}
where ${\cal S}_k$ is  independent of the VEVs
$\vhiggs_u$. This is a very intriguing relation and suggests the
following interpretation.
It is well known that the effect of the FI
term on instantons to modify the instanton moduli space by smoothing
out the singularities corresponding to small instantons. For instance
for a single instanton in $SU(2)$, the centered moduli space is the
singular orbifold ${\mathbb R}^4/{\mathbb Z}_2$, where the radial
coordinate is the instanton size and the $S^3$ is the $SU(2)$
orientation of the instanton. With the FI coupling turned on, the
moduli space is smoothed to the Eguchi-Hanson space. Since the
prepotential involves an integral over the instanton moduli space, the
difference between the prepotential in the commuting and non-commuting
theories loosely speaking involves the contribution from the small instanton
singularities. The interpretation of \eqref{hhh} is then that
small instantons are insensitive to the VEV, as is
clear from the form of the instanton action, and therefore the
contribution from the singularities will be VEV independent. In fact,
the result suggests that the only contribution comes from the
singularity where all the instantons shrink to zero size at the same
point in space. Very
similar ideas have been described in the context of the $\N=4$ theory
in \cite{Partition}. Obviously our discussion here is at best
schematic; however,
our explicit one instanton calculation provides some supporting
evidence where in this case ${\cal S}_1=0$.

\section*{Acknowledgements} 
We would like to thank Diego Bellisai and Chong-Sun Chu for  
discussions. 
G.T. was supported by a PPARC SPG grant 
and the Angelo Della Riccia Foundation.

\end{document}